\documentclass[sigconf,authorversion=true]{acmart}

\usepackage{subcaption}
\usepackage{listings}

\newlength{\twosubht}
\newsavebox{\twosubbox}
\newlength{\foursubht}
\newsavebox{\foursubbox}

\newcounter{hcounter}
\newcommand\heuristic[1]{\refstepcounter{hcounter}\emph{\textbf{Heuristic \arabic{hcounter}}: #1}}

\definecolor{shadecolor}{rgb}{0.9, 0.9, 0.9} 

\newcommand{\gauss}{\mbox{GaussDB}}
\def\naive{na\"ive}
\def\Naive{Na\"ive}

\newcommand{\reqInner}{\mbox{$\delta$}}

\makeatletter
\newcommand{\xleftrightarrow}[2][]{\ext@arrow 3359\leftrightarrowfill@{#1}{#2}}
\newcommand{\xdashrightarrow}[2][]{\ext@arrow 0359\rightarrowfill@@{#1}{#2}}
\newcommand{\xdashleftarrow}[2][]{\ext@arrow 3095\leftarrowfill@@{#1}{#2}}
\newcommand{\xdashleftrightarrow}[2][]{\ext@arrow 3359\leftrightarrowfill@@{#1}{#2}}
\def\rightarrowfill@@{\arrowfill@@\relax\relbar\rightarrow}
\def\leftarrowfill@@{\arrowfill@@\leftarrow\relbar\relax}
\def\leftrightarrowfill@@{\arrowfill@@\leftarrow\relbar\rightarrow}
\def\arrowfill@@#1#2#3#4{%
  $\m@th\thickmuskip0mu\medmuskip\thickmuskip\thinmuskip\thickmuskip
   \relax#4#1
   \xleaders\hbox{$#4#2$}\hfill
   #3$%
}
\makeatother

\AtBeginDocument{%
  }

\copyrightyear{2025} 
\acmYear{2025} 
\setcopyright{cc}
\setcctype{by}
\acmConference[SIGMOD-Companion '25]{Companion of the 2025 International Conference on Management of Data}{June 22--27, 2025}{Berlin, Germany}
\acmBooktitle{Companion of the 2025 International Conference on Management of Data (SIGMOD-Companion '25), June 22--27, 2025, Berlin, Germany}
\acmDOI{10.1145/3722212.3724440}
\acmISBN{979-8-4007-1564-8/2025/06}

\begin{document}

\title{Including Bloom Filters in Bottom-up Optimization}

\author{Tim Zeyl}
\affiliation{%
  \institution{Huawei, Cloud BU}
  \city{Markham}
  \country{Canada}
}
\email{timothy.zeyl@huawei.com}

\author{Qi Cheng}
\affiliation{%
  \institution{Huawei, Cloud BU}
  \city{Markham}
  \country{Canada}
}
\email{qi.cheng1@h-partners.com}

\author{Reza Pournaghi}
\affiliation{%
  \institution{Huawei, Cloud BU}
  \city{Markham}
  \country{Canada}
}
\email{reza.pournaghi1@huawei.com}

\author{Jason Lam}
\affiliation{%
  \institution{Huawei, Cloud BU}
  \city{Markham}
  \country{Canada}
}
\email{jason.lam@huawei.com}

\author{Weicheng Wang}
\affiliation{%
  \institution{Huawei, Cloud BU}
  \city{Markham}
  \country{Canada}
}
\email{ben.wang@huawei.com}

\author{Calvin Wong}
\affiliation{%
  \institution{Huawei, Cloud BU}
  \city{Markham}
  \country{Canada}
}
\email{calvin.wong1@huawei.com}

\author{Chong Chen}
\affiliation{%
  \institution{Huawei, Cloud BU}
  \city{Markham}
  \country{Canada}
}
\email{chongchen@huawei.com}

\author{Per-Ake Larson}
\affiliation{%
  \institution{Huawei, Cloud BU}
  \city{Markham}
  \country{Canada}
}
\email{paul.larson@h-partners.com}

\renewcommand{\shortauthors}{Zeyl et al.}

\begin{abstract}
  
  Bloom filters are used in query processing to perform early data 
  reduction and improve query performance. The optimal query plan may 
  be different when Bloom filters are used, indicating the need for 
  Bloom filter-aware query optimization. 
  To date, Bloom filter-aware query optimization 
  has only been incorporated in a \textit{top-down} query optimizer
  and limited to snowflake queries. 
  In this paper, we show how Bloom filters can be incorporated in a 
  \textit{bottom-up} cost-based query optimizer.   
  We highlight the challenges in limiting 
  optimizer search space expansion, and offer an efficient solution. We show 
  that including Bloom filters in cost-based optimization can lead to better 
  join orders with effective predicate transfer between operators. 
  On a 100 GB instance of the TPC-H database, our approach achieved a 32.8\% 
  further reduction in latency 
  for queries involving Bloom filters, compared to the traditional approach 
  of adding Bloom filters in a separate post-optimization step. Our method applies to all 
  query types, and we provide several heuristics to balance limited increases 
  in optimization time against improved query latency.
\end{abstract}



\maketitle

\section{Introduction}
\label{sec:intro}

Bloom filters and other bit vector filters are used widely in database management 
systems to perform early data reduction~\cite{Bratbergsengen1984,Mackert1986,Chen1993,Chen1997,Das2015}. 
Bloom filters provide an efficient way to probabilistically remove rows early, 
reducing the number of rows participating in further processing and improving query 
performance. Typically, bit vector filters, such as Bloom filters, are
built in hash joins when building the hash join table, and applied to table scans on the 
probe side of that hash join~\cite{Gubner2019,Ding2020,Oracle2024}. 
If the probe side consists of a subtree of operator nodes, a Bloom filter 
can often be pushed down through those operators 
to the table scans. When pushed through an intermediate operator, a Bloom filter 
can also reduce the number of rows participating in that intermediate operator, 
magnifying the improvements observed by using Bloom filters~\cite{Chen1993,Chen1997}.

Given that Bloom filters applied to table scans reduce the number of 
rows produced by a table scan, intentional consideration of this revised row 
estimate during optimization
should lead to potentially better query plans. It was illustrated 
in~\cite{Ding2020} that when Bloom filters are included in a plan, 
the lowest cost join order can be different from the lowest 
cost join order when Bloom filters are absent. Since Bloom filters are 
typically added during post-processing (e.g.,~\cite{Chen1993,Chen1997}), after the optimal query plan 
structure has already been determined, 
the optimal plan that includes Bloom filters may not necessarily be found.

While ~\cite{Ding2020} described how to include Bloom filters 
as a transformation rule in a \textit{top-down} query optimizer~\cite{Graefe1987, Graefe1995}, we do not know 
of any work that describes how to include Bloom filters in a \textit{bottom-up}
optimizer~\cite{Selinger1979,Steinbrunn1997}. We argue that by revising the cardinality estimate for 
scan plans that include Bloom filters, and incorporating a cost model 
for building and applying Bloom filters, a bottom-up optimizer can 
also produce query plans with better join ordering, better join methods, 
and better re-partitioning strategies than simply adding Bloom filters in 
a post process.

\begin{figure}[t!]

    \sbox\twosubbox{%
    \resizebox{\dimexpr.99\columnwidth-1em}{!}{%
        \includegraphics[height=3cm]{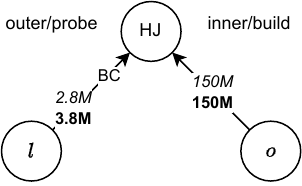}%
        \includegraphics[height=3cm]{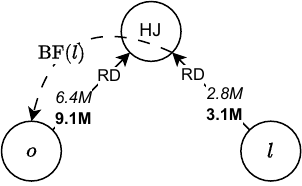}%
    }%
    }
    \setlength{\twosubht}{\ht\twosubbox}

    \centering
    \subcaptionbox{Without BF-CBO\label{fig:q12_bfp1}}{%
        \includegraphics[height=\twosubht]{q12_bfp1}%
    }\quad
    \subcaptionbox{With BF-CBO\label{fig:q12_bfp2}}{%
        \includegraphics[height=\twosubht]{q12_bfp2}%
    }
    \caption{
        The join order for TPC-H query 12 without including Bloom 
        filters during bottom-up CBO (panel a) uses the table \textit{o:orders} 
        with 150M rows as the build side of a hash join (HJ). The 
        \textit{l:lineitem} table with an estimated \textit{2.8M} rows 
        and actual \textbf{3.1M} rows after local predicate filtering is broadcast (BC) to each 
        of the 48 computing threads used in this example. A Bloom 
        filter is not applied during post-processing in this case 
        because the probe side is a foreign key column referencing 
        an \textit{unfiltered} primary key column on the build side; 
        a Bloom filter cannot filter any probe side rows in this case.
        When including
        Bloom filters in CBO (panel b), the join-input ordering is reversed so a Bloom 
        filter can be built on \textit{l:lineitem} and applied to 
        \textit{o:orders}, significantly reducing its estimated row count 
        to \textit{6.4M} rows. The reduced row estimate for \textit{o:orders}
        lowers the cost of this join-input ordering. The planner then 
        selects the lowest cost plan as depicted, which maintains the new 
        join-input ordering and redistributes (RD) both sides. The 
        query runs with 49.2\% lower latency when including Bloom filters 
        in costing.
    }
    \label{fig:q12}
    \Description{Two query join tree diagrams are shown, one obtained without our method, and one with our method.}
\end{figure}

An example is shown in Figure~\ref{fig:q12}, illustrating the join  
produced by our system for TPC-H~\cite{tpchWeb} query 12 with 
and without including Bloom filters in bottom-up 
cost-based optimization (CBO). When Bloom filters are included in costing (BF-CBO), 
the planner explores a plan that applies a Bloom filter to the 
table \textit{orders}, which has a considerably lower estimated row count than 
without the Bloom filter. A better join-input ordering is then selected, resulting in 
far fewer rows needing to participate in the query's join and 
reducing the overall runtime.

At a high level, our method adds additional sub-plans to scan nodes 
where single-column Bloom 
filters can be applied. These Bloom filter scan sub-plans are costed and given 
new cardinality estimates. When evaluating joins between two relations, 
the additional Bloom filter sub-plans are combined with all 
compatible sub-plans from the other relation, and a unique cost and cardinality 
estimate is computed for each combination. This allows a completely different query plan,
relative to a post-processing application of Bloom filters, 
to be built by a bottom-up optimizer.

The optimizer search space is necessarily increased by the inclusion of additional 
Bloom filter sub-plans. Evaluating additional 
sub-plans is a common problem that must be addressed by the optimizer when adding new operators 
to a DBMS. One method we adopt is to impose search-space-limiting heuristics. However,
one nuance that applies uniquely 
to Bloom filter sub-plans, and which we will describe further in Section~\ref{sec:methods}, 
is that their row counts cannot be estimated until the complete set of 
tables on the build side of the join that provides the Bloom filter is known. 
If handled in a \naive{} way, this means the extra Bloom filter sub-plans must be 
maintained with unknown cost until computing the join that provides the required 
Bloom filter build 
relation. This \naive{} approach leads to an explosion 
of the search space and an increase in optimization 
time that is prohibitive.

Our contributions in this paper include 1) introducing a unique property of Bloom filter sub-plans 
in the context of bottom-up CBO that triggers their special handling, and 
2) proposing a two-phased bottom-up approach that minimizes 
the increase in optimizer search space. Our method allows Bloom filters to be considered 
during bottom-up CBO, which enables the optimizer to find different join orderings 
that may provide an opportunity to apply more Bloom filters and provide better predicate transfer.

Next, we describe related work (Section~\ref{sec:related}). In the 
remaining sections of the paper, we describe our method in detail (Section~\ref{sec:methods},
with notations listed in Table~\ref{tab:notation}), 
demonstrate our results (Section~\ref{sec:experiments}), and conclude with 
a discussion (Section~\ref{sec:conclusion}).

\section{Related work}
\label{sec:related}

The idea that Bloom filters can be used for predicate transfer across multiple joins 
is described in~\cite{Yang2023} (Pred-Trans). The authors describe that a predicate on one table ($T_1$) can be 
transferred to a joining table ($T_2$) through a Bloom filter. $T_2$ can realize 
the filtering effects of the predicate on $T_1$ by applying the Bloom filter built from 
the pre-filtered $T_1$. $T_2$ can further transfer that predicate's filtering 
effects to yet another table (e.g., $T_3$) by using another Bloom filter, and so on. 
This idea has its roots in earlier work showing that semi-joins can be used
to reduce the size of relations prior to joins~\cite{Bernstein1981}, and minimize 
data transfer in a distributed context~\cite{Bernstein1981a, Apers1983}. This work indicates 
that finding the best data reduction schedule is its own optimization problem.
The authors of~\cite{Chen1993,Chen1997} extended this line of work by using bit vector filters instead of semi-joins for row reduction.
They showed that the positioning and ordering of bit vector filters applied to 
a fixed execution plan could affect the amount of data reduction observed; again indicating
the need for optimization of a reduction schedule.

In the Pred-Trans paper, Bloom filters were applied 
as a pre-filtering step before any joins were computed. However, the arrangement of Bloom 
filter application was determined heuristically---they arranged the tables 
from small to large, then applied all possible Bloom filters in one forward pass
and one backward pass. The best join order was found independently after the 
input tables had been reduced. This approach was beneficial, but likely missed 
the best arrangement of Bloom filters for optimal predicate transfer, 
as no costed optimization of the reduction schedule was performed.
Our approach differs in that we directly consider the estimated selectivity and cost of 
applying Bloom filters when building the join graph, so that effective 
join orders can be found that inherently consider the predicate transferring ability 
of Bloom filters.

In~\cite{Zhu2017}, the authors study the join order of star-schema 
queries when Bloom filters are built on multiple dimension tables and 
applied to a single fact table. They found that with these Bloom filters 
applied, the query plans were robust to the join order of the multiple 
dimension tables; each join order tested had similar costs. \citeauthor{Ding2020} 
extended this finding to snowflake-schema queries, and demonstrated that the 
best join order for a query can be different when Bloom filters are applied~\cite{Ding2020}. 
They proposed a cost model for Bloom filters and implemented their Bloom filter-aware
optimization as a transformation rule into Microsoft SQL Server~\cite{mssqlserver}, a top-town,
Volcano/Cascades style query optimizer. Their optimizer uses heuristics to detect 
snowflake queries to trigger their Bloom filter transformation rule.

Our work applies to a bottom-up query optimizer, as opposed to a top-down one, 
and applies to all query schemas---not just star-schema or snowflake-schema queries. 
It is general-purpose in its construction because we incorporate the cost and 
cardinality estimate of Bloom filters directly when building each query plan.  

\section{Methods}
\label{sec:methods}

\subsection{Preliminaries and \Naive{} approach}

The basic principle we follow for including Bloom filters in 
bottom-up query optimization is to maintain information about the 
Bloom filter in the nodes to which those Bloom filters 
will be applied.
The well-known process of bottom-up CBO starts by evaluating the cost 
of all supported methods of realizing the base relations in the query, i.e., the different \
ways of scanning the required database tables. Each
of the different ways of accessing those relations can be thought of 
as a \textit{sub-plan}; during bottom-up CBO, a list of sub-plans for each 
relation (including join relations and base relations) is maintained---each 
sub-plan in the relation's \textit{plan-list} represents the lowest cost method with a 
specific set of properties for realizing the corresponding relation.
During the evaluation of a join between two relations, the sub-plans 
from one relation are tested in combination with the sub-plans of the other 
relation, their join cost is computed, and only the lowest cost sub-plans 
for the join relation are kept for the next level of bottom-up CBO 
(where the sub-plans of that join relation may be combined with the sub-plans of other relations).
Higher cost sub-plans are pruned away. This pruning helps to limit 
the search space of sub-plan combinations that the optimizer needs to 
evaluate at higher levels of bottom-up CBO. 

In our approach, we start by adding new Bloom filter sub-plans to base 
relations (or table scans), which include 
additional information about 
the Bloom filter(s) that could be applied to those scans.  All 
Bloom filter information is included on the \textit{apply} side.
These 
additional Bloom filter sub-plans are then included during join evaluation 
when the sub-plans from one relation are combined with those of another. 
The Bloom filter information 
becomes an additional property, and sub-plans with higher costs can be 
pruned according to a common property set~\cite{Lohman1988}. 
Thus, Bloom filter sub-plans are maintained similarly to how interesting join 
orders can be supported~\cite{Simmen1996}.

We consider adding Bloom filter sub-plans on scan nodes only because this ensures that 
the final plan will push down Bloom filters as far as possible to those 
scan nodes. This means we must build our Bloom filters using hash keys derived from values in
single columns, rather than supporting hash keys based on values across multiple columns. 
So, when a join consists of multiple join columns, instead of building
a multi-column Bloom filter we plan for, and build, separate single-column Bloom filters.
Additionally, we only consider building Bloom filters on the build side 
of hash join nodes. While building in other nodes could be supported, 
following this convention allows us to ensure Bloom filters will be 
fully built before they are used on the probe side.

The Bloom filter property we add to scan node sub-plans can propagate up through joins. It 
may be present in the sub-plans created for join relations,
or it can be removed if the joined relation \textit{resolves} the Bloom 
filter---that is, if the joined relation provides the required build side of the Bloom filter sub-plan.
The Bloom filter property differs from other properties, like sort order, in one 
important sense, however. 
That is, the cardinality (or estimated row count) of a scan with a Bloom filter applied depends on the 
set of relations involved on the build side of the hash join that 
creates the Bloom filter. 

\begin{table}[tb]
    \caption{Table of notations}
    \centering
    \begin{tabular}{|l|p{0.6\linewidth}|}
        \hline
        \textbf{Notation} & \textbf{Description} \\ \hline
        $R_n$ & A base relation, or table, identified by $n$. \\ \hline
        $|R_n|$ & The row count, or cardinality, of relation $R_n$. \\ \hline
        $(R_0,R_1)$ & A joined relation involving a join between relations $R_0$ and $R_1$. \\ \hline
        $R_0 {\ltimes} (R_1,R_2)$ & $R_0$ semi-join $(R_1,R_2)$. \\ \hline
        $R_0 \hat{\ltimes} (R_1,R_2)$ & The result of a bloom filter built from a column of $(R_1,R_2)$ and applied to $R_0$. 
            This can be thought of as an approximate semi-join. \\  \hline
        $R_0 \xdashleftarrow{\textrm{BF}(R_1)} (R_1,R_2)$ & Alternate notation for $R_0 \hat{\ltimes} (R_1,R_2)$ used in diagrams. 
            The Bloom filter build column comes from $R_1$. \\ \hline
        $\delta=\{R_a,...,R_z\}$ & The set of base relations required to appear on the build-side of a hash join with a Bloom filter sub-plan. \\ \hline
        $\Delta = [\delta_0,...,\delta_n]$ & A list of possible $\delta$'s that can be used to create different Bloom filter sub-plans on a base relation. \\ \hline
    \end{tabular}
    \label{tab:notation}
\end{table}

\begin{figure}[t!]

    \sbox\twosubbox{%
    \resizebox{\dimexpr.99\columnwidth-1em}{!}{%
        \includegraphics[height=3cm]{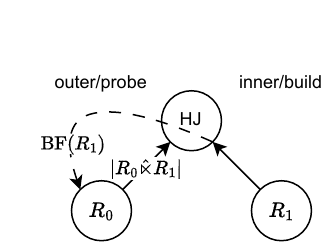}%
        \includegraphics[height=3cm]{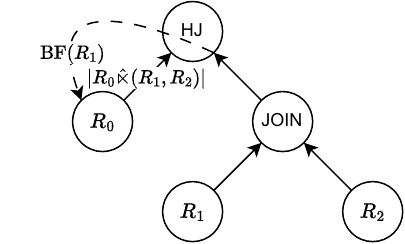}%
    }%
    }
    \setlength{\twosubht}{\ht\twosubbox}

    \centering
    \subcaptionbox{\reqInner{}$ = \{R_1\}$\label{fig:bf_req_inner_a}}{%
        \includegraphics[height=\twosubht]{bf_req_inner_demo_R1inner}%
    }\quad
    \subcaptionbox{\reqInner{}$ = \{R_1,R_2\}$\label{fig:bf_req_inner_b}}{%
        \includegraphics[height=\twosubht]{bf_req_inner_demo_R1R2inner}%
    }
    \caption{
        The cardinality of a scan on $R_0$ after applying a Bloom filter built from 
        a column of relation $R_1$ (denoted by the dashed arrow labeled BF($R_1$)) depends on the set of relations 
        on the build side of the hash join in which that Bloom filter is created. 
        The cardinality of $R_0$ is $|R_0 \hat{\ltimes} R_1|$
        when the build side consists solely of relation $R_1$, as in panel (\subref{fig:bf_req_inner_a}). So in 
        this case, the Bloom filter sub-plan for $R_0$ has property \reqInner{} = $\{R_1\}$; 
        it requires the relation $R_1$ be joined to it for the Bloom 
        filter to be resolved.         
        When the build side includes both $R_1$ and $R_2$ (panel \subref{fig:bf_req_inner_b}), the 
        cardinality of $R_0$ after the Bloom filter is applied becomes 
        $|R_0 \hat{\ltimes} (R_1,R_2)|$, which may be lower than that in panel (\subref{fig:bf_req_inner_a}). 
        In this case, the Bloom filter sub-plan
        for $R_0$ has property \reqInner{} = $\{R_1,R_2\}$; it requires
        both relations $R_1$ and $R_2$ to be joined to it for its Bloom 
        filter to be resolved. Note that the build (inner) side of the hash join is depicted 
        on the right side in our convention.
    }
    \label{fig:bf_req_inner}
    \Description{Two query join tree diagrams are shown demonstrating the difference in cardinality 
        estimate depending on which relations appear on the right side of a hash join.}
\end{figure}

To explain this dependency, we first note that the 
cardinality of a relation $R_0$ with a 
Bloom filter built from a single relation $R_1$ (denoted $|R_0 \hat{\ltimes} R_1|$ and 
shown in Figure~\ref{fig:bf_req_inner_a}) can 
be estimated as the cardinality of 
the semi-join of $R_0$ with $R_1$ (denoted $|R_0 \ltimes R_1|$), 
plus some Bloom filter false positive rate. In other words, 
$|R_0 \hat{\ltimes} R_1| \geq |R_0 \ltimes R_1|$, where equality 
occurs if the false positive rate is 0. Then, we note that 
$|R_0 \ltimes R_1| \ge |R_0 \ltimes (R_1, R_2,..., R_n)|$ if the 
semi-join clause is between $R_0$ and $R_1$, because any joins that $R_1$ 
has with other relations before joining with $R_0$ may remove some distinct elements of the joining 
$R_1$ column, and reduce the number of elements in a Bloom filter 
built on $R_1$. This, in turn, would reduce the number of rows coming 
out of a scan that applies a Bloom filter on $R_0$. 
So, to estimate the cardinality, and therefore cost, of a 
Bloom filter sub-plan 
we must know the set of relations that appear on the build side of the 
hash join. As shown in Figure~\ref{fig:bf_req_inner}, the cardinality of 
$R_0$ with a Bloom filter built from $R_1$ is different when $R_1$ is 
first joined to $R_2$.

This dependency poses a problem for a bottom-up optimizer, where the 
set of relations appearing on 
the build side of a join is not generally known \textit{a priori}. Because plans 
are built bottom up, the cardinality of a Bloom filter sub-plan on $R_0$ cannot be 
known until the optimizer knows the set of relations to which that sub-plan will be joined.
A \naive{} 
solution may maintain several uncosted sub-plans with 
\textit{unresolved} Bloom filter information. These uncosted, 
unresolved sub-plans 
would inevitably be combined with relations that do not provide 
the build side of the Bloom filter and, while uncosted, these sub-plans 
cannot be pruned, so the number of sub-plans that need to be maintained would 
grow exponentially with each join that does not \textit{resolve} 
the Bloom filter. A Bloom filter sub-plan can only be fully costed when 
the Bloom filter becomes resolved, i.e., when evaluating 
a join where the requisite relation appears on the build side. It is 
only then that the cardinality of Bloom filter sub-plans can be 
estimated, a necessarily recursive process in which the sub-plan is 
traversed to the leaf table scan whose cardinality and cost must be computed
with respect to the now-known set of relations on the hash join build side. 
The computed cardinality of the leaf table scan, in turn, influences 
the cardinality and cost of any intermediate plan nodes back up to the
resolution node. We found a \naive{} 
approach like this led to prohibitive optimization times: 
for example, a 3-table join query took 28 ms to optimize, 
a 4-table join query took 375 ms, and a 5-table join query took 56 s. A 6-table 
join did not finish its optimization in more than 30 min, and we did not wait longer.

\subsection{Our approach: BF-CBO}

The key to preventing this exponential growth in the search space is to delay 
planning until pruning is possible. That way, numerous 
uncosted plans need not be maintained. We achieve this 
property of delaying planning until pruning is possible with a 
two-phase bottom-up optimization approach. At 
a high level, this involves the following steps, which we will describe in 
detail in the following sections.

\begin{enumerate}
\item \emph{Marking Bloom filter candidates}. We first identify which base 
tables are suitable candidates for applying a Bloom filter and attach the 
required Bloom filter information to those tables.
\item \emph{First bottom-up phase}. We compute a first 
bottom-up pass in which all the valid join combinations are decided,
and all sets of relations that appear 
on the build-side of a join with Bloom filter 
candidate relations are identified (we can denote one such set as $\delta=\{R_a,R_b,...,R_z\}$, 
the required build-side relations for a Bloom filter sub-plan to be resolved).
\item \emph{Costing Bloom filter sub-plans}. We create new Bloom filter scan sub-plans 
on base tables and estimate their cardinality and cost according
to \reqInner{}.
\item \emph{Second bottom-up phase}. We compute a second bottom-up pass in which 
all intermediate sub-plans are fully costed and planned, with any Bloom filter sub-plans adherent to 
join order restrictions set out by the assumptions inherent in \reqInner{}.
\end{enumerate}

\subsection{Marking Bloom filter candidates}
\label{sec:methods_marking_bfcs}

Our solution begins after the optimizer has estimated the cardinality 
of each base relation (i.e., table scan) and added sub-plans for the 
scan of those base relations, but before the optimizer has created any 
join sub-plans or evaluated the cost of any join combinations.
We first identify all \textit{Bloom filter 
candidates} that can be applied to base relations based on the join clauses 
in the query. 

A Bloom filter candidate includes information identifying the join clause, the
Bloom filter build table and column, as well as the apply column, and it 
is attached as additional information to the base relation to which 
the associated Bloom filter can be applied (no information is attached to 
the build-side relation). Multiple Bloom filter 
candidates from different join clauses can be attached to the same 
relation. Each Bloom filter candidate has an initially empty list, $\Delta = [~]$, of 
required build-side relation sets that become populated during the 
first bottom-up pass (e.g., $\Delta = [\delta_0,\delta_1,...,\delta_n]$), 
and which identifies all valid Bloom filter 
scan sub-plans. Bloom filter candidates should be thought of as being a property of 
the relation to which a Bloom filter might be applied, rather than as a property 
of the many sub-plans that can realize that relation.

We use some 
heuristics to help us 
limit the total number of Bloom filter candidates, which can 
reduce the number of additional sub-plans that must be evaluated during CBO. These heuristics 
are implemented throughout our method, so we'll enumerate them in the text and summarize them in Section~\ref{sec:heuristics}.
First, for a pair of relations in a join clause, we only include a Bloom filter candidate
on the larger of the two tables (\textbf{Heuristic~\ref{h:larger_rel}}), as it is often more likely 
the Bloom filter has a greater filtering capacity in this configuration. If we have a multi-way equivalence clause, 
then we only consider building a Bloom filter from the smallest table 
and applying it to the larger tables. Second, if the estimated number of rows on 
the apply-side table is below a threshold, it may not be worthwhile to apply 
a Bloom filter anyway, so we do not include a Bloom filter candidate in that case (\textbf{Heuristic~\ref{h:apply_threshold}}).
Further heuristics are applied during bottom-up optimization. 

We also restrict applying any Bloom filter candidates 
if the build and apply column will cross a full outer join or an anti join; applying 
a Bloom filter in these cases could yield incorrect results, so this restriction 
is not considered a heuristic. Similarly, 
if a Bloom filter candidate's build and apply column were to cross a left outer join, 
we must restrict the apply column from being on the row-preserve (left) side,
as that would also yield incorrect results.

We will now introduce a running example for each step of our method.

\lstset{
    escapeinside={(*@}{@*)}
}

\begin{example}\noindent \textbf{Marking Bloom filter candidates}.
    \label{ex:1}
    \noindent Consider the example query:

    \begin{lstlisting}[language=SQL]
    SELECT *
    FROM (*@$t1,t2,t3$@*)
    WHERE (*@$t1.c2 = t2.c1$@*)
        AND (*@$t2.c2 = t3.c1$@*)
        AND (*@$t2.c3 < 100$@*);
    \end{lstlisting}

    \noindent with the following estimated base relation cardinalities:

    \smallskip

    \begin{tabular}{|l|r|}
        \hline
        \textbf{Relation} & \textbf{row estimate} \\ \hline
        $t1$ & 600M \\ \hline
        $t2$ WHERE $t2.c3$ < 100 & 807K \\ \hline 
        $t3$ & 1M \\
        \hline
    \end{tabular}

    \smallskip

    \noindent and $t2.c2$ is a foreign key of $t3.c1$.

    For each hashable join clause we may place one Bloom filter candidate (BFC). So for $t1.c2 = t2.c1$,
    we place a BFC on $t1$ because it has a larger cardinality than $t2$ (Heuristic~\ref{h:larger_rel}). Similarly, for 
    $t2.c2 = t3.c1$ we place a BFC on $t3$ because it has a larger cardinality than $t2$. In summary, 
    we have the following BFCs:
    \begin{itemize}
        \item $t1.\textrm{bfc}_1: a=t1.c2, b=t2.c1, \Delta=[~]$
        \item $t3.\textrm{bfc}_1: a=t3.c1, b=t2.c2, \Delta=[~]$
    \end{itemize}
    where $a$ records the apply-side relation and column, and $b$ records the build-side relation and column.

\end{example}

\subsection{First bottom-up phase}
\label{sec:methods_phase1}

In the first bottom-up pass, we simulate the process of 
combining relations as in normal bottom-up CBO. However, instead 
of costing any sub-plans, we only populate the list of 
\reqInner{} relation sets, $\Delta$, that are observed during 
this process. For example, for a join involving three relations, 
$R_0,R_1,R_2$, if we have a Bloom filter candidate applied to $R_0$ 
built from a column of $R_1$, then during this process we may observe 
these two relation sets:
$\delta_0=\{R_1\}$ and $\delta_1=\{R_1,R_2\}$---as indicated by the join 
orders in Figure~\ref{fig:bf_req_inner}; so the list of possible 
relation sets for that Bloom filter candidate on $R_0$ would be $\Delta = [\delta_0,\delta_1]$. 

During the first bottom-up pass, we can also prune any \reqInner{}s where the Bloom 
filter candidate join clause consists of a foreign key on the apply side 
referencing a lossless primary key on the build side (\textbf{Heuristic~\ref{h:fk_pk}}). When the primary 
key column is unfiltered and is the column used to build the Bloom filter, 
then we know that it will not filter any rows on the apply side, so 
we need not create a Bloom filter scan sub-plan for that scenario. This heuristic 
is implemented here because we can only determine if the primary key is lossless 
with respect to this sub-plan
once we know the complete set of relations, $\delta$, to appear on the build side 
of the join.

\begin{example}\noindent \textbf{First bottom-up phase}.
    \label{ex:2}
    \noindent Continuing from Example~\ref{ex:1}, during the first bottom-up phase, we'd observe the 
    following ordered join combinations (grouped by the join relations they create), and would populate $\Delta$ for each 
    BFC, if possible. In each case, we defer computing any join sub-plans.

    \paragraph{Join Relation: $(t1,t2)$.}
    \begin{itemize}
        \item $t1$ JOIN $t2$: Here, $t1$ is the outer relation and $t2$ is the inner relation. 
        $t1$ has a BFC, namely $t1.\textrm{bfc}_1$, and the inner relation ($t2$) supplies the required build column. 
        So we populate $\Delta$ with the inner relations observed for this join pair (i.e., $\delta=\{t2\}$). 
        The updated BFC is
        \begin{equation*}
        t1.\textrm{bfc}_1: a=t1.c2, b=t2.c1, \Delta=[\{t2\}]
        \end{equation*}
        \item $t2$ JOIN $t1$: Here, $t2$ is the outer relation and $t1$ is the inner relation, so this join pair cannot supply 
        the build column for $t1.\textrm{bfc}_1$ because the Bloom filter must be built on the 
        inner (build) side of a hash join. There is nothing to do for this join pair.
    \end{itemize}
    \paragraph{Join Relation: $(t2,t3)$.}
    \begin{itemize}
        \item $t2$ JOIN $t3$: Here, $t3$ has a BFC ($t3.\textrm{bfc}_1$), but it is on the inner 
        side of the join, so its build column cannot be resolved here. There is nothing to do for this join pair.
        \item $t3$ JOIN $t2$: Here, $t3$ is the outer relation and the inner is $t2$ which supplies the required build column. 
        So we populate $\Delta$ as follows:
        \begin{equation*}
        t3.\textrm{bfc}_1: a=t3.c1, b=t2.c2, \Delta=[\{t2\}]
        \end{equation*}
    \end{itemize}
    \paragraph{Join Relation: $(t1,t2,t3)$}
    \begin{itemize}
        \item $(t1,t2)$ JOIN $t3$: Here, the inner relation is $t3$, and it does not supply the build column of $t1.\textrm{bfc}_1$, so there is nothing to do.
        \item $t3$ JOIN $(t1,t2)$: Here, the inner relation is $(t1,t2)$ which supplies the build column of $t3.\textrm{bfc}_1$. We update its $\Delta$ as follows:
        \begin{equation*}
        t3.\textrm{bfc}_1: a=t3.c1, b=t2.c2, \Delta=[\{t2\},\{t1,t2\}]
        \end{equation*}
        \item $(t2,t3)$ JOIN $t1$: Here, the inner relation, $t1$, does not supply the build relation for the outer BFC ($t3.\textrm{bfc}_1$), so there is nothing to do.
        \item $t1$ JOIN $(t2,t3)$: Here, the inner relation is $(t2,t3)$ which supplies the build column of $t1.\textrm{bfc}_1$. We update its $\Delta$ as follows:
        \begin{equation*}
        t1.\textrm{bfc}_1: a=t1.c2, b=t2.c1, \Delta=[\{t2\}, \{t2,t3\}]
        \end{equation*}
    \end{itemize}

    In this example, we did not encounter any join pairs where the Bloom 
    filter candidate join clause consists of a foreign key on the apply side 
    referencing a lossless primary key on the build side. So, we did not prune any 
    potential $\delta$s due to Heuristic~\ref{h:fk_pk}.

\end{example}

\subsection{Costing Bloom filter sub-plans}
\label{sec:methods_costing_subplans}

After the first bottom-up pass, each Bloom filter candidate should have a list of valid 
\reqInner{}s. For each of those 
\reqInner{}s, we create a scan sub-plan that includes the 
application of a Bloom filter during scanning. We compute the 
cardinality of this \textit{Bloom filter sub-plan} using 
the estimated selectivity of a 
semi-join of this relation and those in \reqInner{}, 
plus the estimated Bloom filter false positive rate. 
For example, 
we may create one Bloom filter sub-plan for the scan of $R_0$ with $\delta = \{R_1\}$ 
with a cardinality estimate of $|R_0 \hat{\ltimes} (R_1)|$. We may 
create another Bloom filter sub-plan for the scan of $R_0$ with $\delta = \{R_1, R_2\}$ 
with a cardinality estimate of $|R_0 \hat{\ltimes} (R_1, R_2)|$, which may 
have fewer estimated rows but also require more relations on the build 
side of the hash join on which the Bloom filter will be built.

After creating these new Bloom filter sub-plans, we attempt to add them 
to the base relation's list of lowest-cost sub-plans (i.e., the relation's \textit{plan-list}). 
During this process,
the new Bloom filter sub-plans
can be pruned against one another based on the property of 
\reqInner{}, as follows. 
If a new sub-plan requires more relations 
in \reqInner{} than pre-existing sub-plans in the relation's plan-list, 
but it has fewer rows, then it will be kept as an interesting option, 
regardless of cost. If the new sub-plan requires more relations 
in \reqInner{}, but it \textit{does not} have fewer rows, then 
we know the extra required relations in \reqInner{} 
did not provide more filtering capacity and the new sub-plan can be immediately 
pruned. This immediate pruning helps to limit the search space explored
throughout bottom-up optimization.

If multiple Bloom filters are candidates on the same 
relation (originating from different join clauses), then we create 
Bloom filter scan plans with all 
possible Bloom filter candidates applied as opposed to testing out 
sub-plans with various subsets of Bloom filter candidates applied.
In other words, we apply all 
valid candidate Bloom filters on a base table \textit{simultaneously}, as an 
additional heuristic to limit the search space (\textbf{Heuristic~\ref{h:all_BFCs}}). We 
do, however, allow for various combinations of $\delta$s when creating 
new Bloom filter scan sub-plans.

The Bloom filter false positive 
rate can be derived from the number of bits in the Bloom filter array 
and the number of hash functions used in the Bloom filter. The number 
of bits in the Bloom filter is determined through 
an upper bound estimate of the number of distinct values on the 
Bloom filter build side. The number of hash functions is fixed at two 
for performance reasons. 

We remove any sub-plans whose estimated Bloom filter size is beyond a 
threshold (\textbf{Heuristic~\ref{h:BF_too_big}}). The purpose of this restriction is 
to limit the size of created Bloom filters so that they can mostly be 
accommodated by the L2 cache. If a Bloom filter spills beyond this, 
there will be slowdowns in accessing and probing that Bloom filter, reducing 
its benefits. We also remove any Bloom filter sub-plans whose estimated
selectivity (excluding false positives) is lower than a threshold  (\textbf{Heuristic~\ref{h:selec_threshold}}). In 
this way we only retain Bloom filter candidates that are likely to have 
a large enough filtering capacity. 

We model the cost of applying the Bloom filter as a constant value ($k$)
times the number of rows to be filtered to represent the cost of evaluating 
the Bloom filter hash functions for each row. $k$ is set to be smaller 
than the cost per row of a regular hash table lookup. We also provide a mechanism 
to account for the cost of building each Bloom filter, but in practice 
we found this cost to be negligible, so it is set to zero in our cost model.

\begin{example}\noindent \textbf{Costing Bloom filter sub-plans}.
    \label{ex:3}
    \noindent Continuing from Example~\ref{ex:2}, after the first bottom-up phase, we have the 
    following Bloom filter candidates:

    \begin{itemize}
        \item $t1.\textrm{bfc}_1$:
        \begin{itemize}
            \item $a=t1.c2, b=t2.c1$
            \item $\Delta=[\{t2\},\{t2,t3\}]$
        \end{itemize}
        \item $t3.\textrm{bfc}_1$:
        \begin{itemize}
            \item $a=t3.c1, b=t2.c2$, 
            \item $\Delta=[\{t2\},\{t1,t2\}]$
        \end{itemize}
    \end{itemize}

    
    For the scan of $t1$ we create two Bloom filter sub-plans, each with 
    a single Bloom filter applied, but in each sub-plan that Bloom filter
    has a unique $\delta$, one for each $\Delta$ in the Bloom filter candidate
    above. We include the cardinality for each of the sub-plans as follows:

    \begin{itemize}
        \item $t1.\textrm{bf-subplan}_0$:
        \begin{itemize}
            \item $\textrm{bfs}=[(a=t1.c2, b=t2.c1, \delta=\{t2\})]$
            \item $\textrm{rows}=|t1 \hat{\ltimes} t2|$
        \end{itemize}
        \item $t1.\textrm{bf-subplan}_1$:
        \begin{itemize}
            \item $\textrm{bfs}=[(a=t1.c2, b=t2.c1, \delta=\{t2,t3\})]$
            \item $\textrm{rows}=|t1 \hat{\ltimes} (t2,t3)|$
        \end{itemize}
    \end{itemize}

    \noindent Now, suppose the Bloom filters in both of these sub-plans yield the same 
    estimated cardinality; for example, $|t1 \hat{\ltimes} t2| = |t1 \hat{\ltimes} (t2,t3)| = 22M$.
    In this case, the optimizer believes there is no added benefit in 
    first joining $t2$ to $t3$ when building the Bloom filter. This follows 
    from the fact that there is no local predicate on $t3$ that could be 
    transferred through the join of $t2$ and $t3$ to $t1$. 

    Next, we compute the cost for each sub-plan and try to add each sub-plan 
    to the base relation's plan-list. $t1$'s plan-list should 
    have at least one pre-existing non-Bloom filter costed sub-plan with a row 
    estimate of 600 million. Then, for each Bloom filter sub-plan, we add extra cost 
    for applying the Bloom filter to each input row of $t1$ as 
    \begin{equation*}
    \textrm{extra cost} = k*600M.
    \end{equation*}

    \noindent So, the cost of each Bloom filter sub-plan is equal. The first sub-plan, $t1.\textrm{bf-subplan}_0$,
    will be accepted in $t1$'s plan-list because it lowers the row count to 22 million (compared to 600 million), but the second 
    sub-plan, $t1.\textrm{bf-subplan}_1$,
    will be rejected since it has the same row count (22 million) and cost as $t1.\textrm{bf-subplan}_0$ and 
    requires an additional relation on the build side 
    of the join.


    For the scan of $t3$, we similarly have two Bloom filter sub-plans:

    \begin{itemize}
        \item $t3.\textrm{bf-subplan}_0$
        \begin{itemize}
            \item $\textrm{bfs}=[(a=t3.c1, b=t2.c2, \delta=\{t2\})]$
            \item $\textrm{rows}=|t3 \hat{\ltimes} t2|$
        \end{itemize}
        \item $t3.\textrm{bf-subplan}_1$
        \begin{itemize}
            \item $\textrm{bfs}=[(a=t3.c1, b=t2.c2, \delta=\{t1,t2\})]$
            \item $\textrm{rows}=|t3 \hat{\ltimes} (t1,t2)|$
        \end{itemize}
    \end{itemize}

    \noindent Now, in this example, the selectivity of the semi-join $t3 \ltimes t2$ 
    is 0.77, which is beyond our threshold, so $t3.\textrm{bf-subplan}_0$
    is rejected by Heuristic~\ref{h:selec_threshold}.
    However, the selectivity of semi-join $t3 \ltimes (t2,t1)$ is 0.006, 
    yielding a cardinality estimate for sub-plan $t3.\textrm{bf-subplan}_1$ 
    of 36 thousand rows, much lower than existing sub-plan row estimates of 1 million for 
    $t3$. So, this sub-plan is accepted in $t3$'s plan-list. The extra cost 
    is $k*1M$.
    
\end{example}

\subsection{Second bottom-up phase}

\begin{figure*}[t!]

    \sbox\foursubbox{%
    \resizebox{\dimexpr0.98\textwidth-0.2em}{!}{%
        \includegraphics[height=3cm]{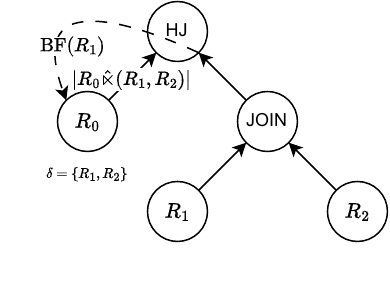}%
        \includegraphics[height=3cm]{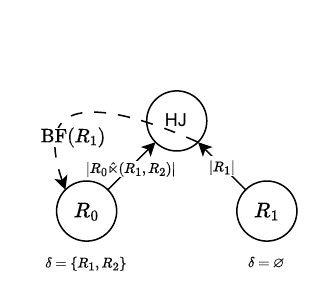}%
        \includegraphics[height=3cm]{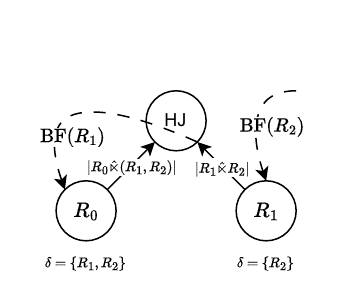}%
        \includegraphics[height=3cm]{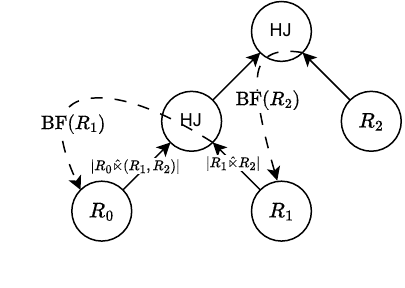}%
    }%
    }
    \setlength{\foursubht}{\ht\foursubbox}

    \centering
    \subcaptionbox{$R_0$ joined with $(R_1, R_2)$\label{fig:rule_demo_a}}{%
        \includegraphics[height=\foursubht]{bf_restrictions_a}%
    }\hfill
    \subcaptionbox{Illegal sub-plan join\label{fig:rule_demo_d}}{%
        \includegraphics[height=\foursubht]{bf_restrictions_d}%
    }\hfill
    \subcaptionbox{Allowed sub-plan join\label{fig:rule_demo_b}}{%
        \includegraphics[height=\foursubht]{bf_restrictions_b}%
    }\hfill
    \subcaptionbox{Resolved sub-plan for (c)\label{fig:rule_demo_c}}{%
        \includegraphics[height=\foursubht]{bf_restrictions_c}%
    }
    \caption{
        During the first bottom-up pass, the join of $R_0$ to $(R_1,R_2)$ 
        may have been observed, as depicted in panel (\subref{fig:rule_demo_a}). 
        Our approach creates a sub-plan for the scan of $R_0$ with a Bloom
        filter built from $R_1$ and an estimated cardinality of $|R_0 \hat{\ltimes} (R_1,R_2)|$. 
        The sub-plan has property \reqInner{}$=\{R_1,R_2\}$ because both
        $R_1$ and $R_2$ are required on the build side of the hash join (HJ) 
        for the cardinality of that sub-plan 
        to be accurate. During the second bottom-up pass, we forbid 
        the join between sub-plans as depicted in panel (\subref{fig:rule_demo_d})
        because $R_2$ does not appear on the build side of the HJ, violating 
        the cardinality assumptions on the sub-plan of $R_0$. Panel (\subref{fig:rule_demo_b})
        shows an allowed join because, even though $R_2$ does not appear 
        on the build side of the HJ, $R_0$ is being joined with a sub-plan 
        that requires a BF built from $R_2$ (i.e., the sub-plan of $R_1$ has property $\reqInner{}=\{R_2\}$). So, any filtering that 
        $R_2$ does on $R_1$ will effectively be transferred to $R_0$ through 
        the Bloom filter applied to $R_1$ (i.e., BF($R_2$)).         
        The incomplete sub-plan in panel (\subref{fig:rule_demo_b}) may 
        be completed in the next level of bottom-up optimization by a HJ with $R_2$ as shown in panel (\subref{fig:rule_demo_c}),
        which we consider to be equivalent to the join order in panel (\subref{fig:rule_demo_a})
        (ignoring any Bloom filter false positives).
    }
    \label{fig:rule_demo}
    \Description{Four sub-figures are depicted, each showing the cardinality estimates in different join configurations. 
        In panel a, R0 is joined to a subtree containing R1 and R2 on the inner (right) side of hash join.
        In panel b, the inner relation is R1, whose sub-plan does not require R2. 
        In panel c, the inner relation is R1, and R1s sub-plan does require a BF applied from R2.
        In panel d, R2 is joined to a copy of the diagram in panel c, above a top level hash join.}
\end{figure*}

At the beginning of the second bottom-up pass, there now exists fully 
costed sub-plans for accessing every base table in the query. 
Some of these sub-plans may include the application of a Bloom filter (so-called \textit{Bloom filter sub-plans})
and contain information about that Bloom filter, namely the build 
and apply columns, as well as the set of required \reqInner{} 
relations. Since all Bloom filter sub-plans now have a cardinality estimate and are 
fully costed, bottom-up optimization can proceed as usual, subject to some 
additional constraints.

First, when joining a Bloom filter sub-plan to another sub-plan that provides 
the build relation for the Bloom filter, the join method for that pair 
of sub-plans must be a hash join, and the Bloom filter sub-plan must be on 
the outer (probe) side. Other join types will be satisfied by non-Bloom filter sub-plans.
Second, if the other sub-plan provides any relation 
listed in the Bloom filter sub-plan's \reqInner{}, then 
the join method must be a hash join because the cardinality of the 
Bloom filter sub-plan assumes all the relations in \reqInner{} 
will be on the inner (build) side of a hash join. Consequently, the inner 
side of the join must provide \textit{all} relations in 
\reqInner{}, with one exception: the inner side need not 
provide \textit{all} relations in 
\reqInner{} if it is also a Bloom filter sub-plan whose 
\reqInner{} match the outstanding relations in the outer's \reqInner{}.
Figure~\ref{fig:rule_demo} illustrates why we allow this exception.

These additional constraints limit how we combine Bloom filter 
sub-plans, but they do not preclude the evaluation of all other non-Bloom filter 
sub-plans. So, during this second bottom-up phase, we are evaluating 
more combinations of sub-plans than before: all the original non-Bloom filter 
sub-plans plus newly added Bloom filter sub-plans. Despite adhering to the 
property of delaying planning until sub-plans can be pruned, the search 
space is still expanded, and as such, we expect planning time to be increased.
We discuss several additional heuristics to combat this increase in Section~\ref{sec:heuristics}.

When a Bloom filter sub-plan from a relation $R_0$ is joined to a 
sub-plan from another relation $R_1$ that resolves 
\textit{all} Bloom filters in that sub-plan, the cost of the corresponding hash 
join for this sub-plan pair is computed. The cardinality estimate 
simply becomes the original cardinality estimate for the joined relation $(R_0,R_1)$
because all Bloom filters have been resolved in the joined sub-plan. 
Accordingly, the Bloom filter information is 
removed from the joined sub-plan so that it can compete against any other 
non-Bloom filter sub-plans in the joined relation $(R_0,R_1)$. 

When a Bloom filter sub-plan from relation $R_0$ is joined to a sub-plan 
from another relation $R_2$ that does \textit{not} resolve all Bloom filters in 
that sub-plan, then the Bloom filter information from any unresolved 
Bloom filters is retained in the joined sub-plan. For example, if we have 
a sub-plan of $R_0$ with required build-side relations $\delta = \{R_1\}$, 
when we join that sub-plan to a sub-plan of $R_2$, the Bloom filter will not be resolved, so the 
new sub-plan of joined relation $(R_0,R_2)$ will retain the property 
$\delta = \{R_1\}$. Its cardinality will typically be lower than other non-Bloom filter 
sub-plans in the plan-list for $(R_0,R_2)$.

\begin{figure}[t!]

    \sbox\twosubbox{%
    \resizebox{\dimexpr.99\columnwidth-1em}{!}{%
        \includegraphics[height=3cm]{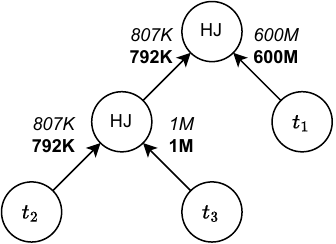}%
        \includegraphics[height=3cm]{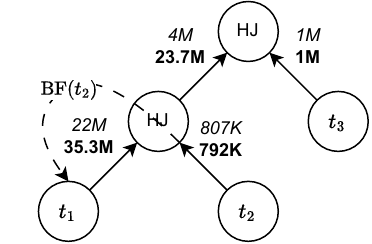}%
    }%
    }
    \setlength{\twosubht}{\ht\twosubbox}

    \centering
    \subcaptionbox{BF-Post\label{fig:running_ex_bfp1}}{%
        \includegraphics[height=\twosubht]{running_ex_bfp1}%
    }\quad
    \subcaptionbox{BF-CBO\label{fig:running_ex_bfp2}}{%
        \includegraphics[height=\twosubht]{running_ex_bfp2}%
    }
    \caption{
        Join orders obtained for the running example in Section~\ref{sec:methods}. 
        Observed input row counts are shown in bold on the left (outer/probe) and 
        right (inner/build) sides of each join; the planner's row estimates are italicized. 
        Post-processing application of Bloom filters (panel \subref{fig:running_ex_bfp1}), 
        does not apply any Bloom filters. BF-CBO (panel \subref{fig:running_ex_bfp2})
        modifies the join order to apply a Bloom filter on $t1$, significantly 
        reducing the observed input row counts of each join.
    }
    \label{fig:running_ex}
    \Description{Two subfigures are depicted, each showing a join tree with different table ordering.}
\end{figure}

\begin{example}\noindent\textbf{Second bottom-up phase}.
    \label{ex:4}
    \noindent Continuing from Example~\ref{ex:3}, after Bloom filter costing, we have the following 
    Bloom filter sub-plans included in the plan-list for each of the base relations 
    $t1$ and $t3$. 
    \begin{itemize}
        \item $t1.\textrm{bf-subplan}_0$:
        \begin{itemize}
            \item $\textrm{bfs}=[(a=t1.c2, b=t2.c1, \delta=\{t2\})]$
            \item $\textrm{rows}=22M$
        \end{itemize}
        \item $t3.\textrm{bf-subplan}_1$
        \begin{itemize}
            \item $\textrm{bfs}=[(a=t3.c1, b=t2.c2, \delta=\{t1,t2\})]$
            \item $\textrm{rows}=36K$
        \end{itemize}
    \end{itemize}

    \noindent We also have all existing non-Bloom filter sub-plans in the respective plan-lists for all base relations.
    
    Next, we'd evaluate joining all combinations of sub-plans from the base 
    relations, building a costed plan bottom-up. We'd observe the same 
    ordered join pairs from the first bottom-up phase, but this time we'd
    evaluate the cost of all join types, namely nest loop join, merge join, and hash join 
    for all sub-plans, including the new Bloom filter sub-plans.

    \paragraph{Join Relation: $(t1,t2)$.}
    \begin{itemize}
        \item $t1$ JOIN $t2$: Here, the required Bloom filter in sub-plan \linebreak $t1.\textrm{bf-subplan}_0$ can be 
        resolved by the inner relation $t2$. We compute the cost of the corresponding hash join and allow 
        it to compete with the existing sub-plans in the plan-list for relation $(t1,t2)$. It is accepted 
        and removes several other higher-cost sub-plans from the plan-list. Since the joined sub-plan no longer 
        requires any Bloom filters, the set of required build-side relations becomes null (i.e., $\delta=\varnothing$).
        \item $t2$ JOIN $t1$: Here, the required Bloom filter in sub-plan \linebreak $t1.\textrm{bf-subplan}_0$ cannot be resolved because the Bloom filter build relation, $t2$, is on the outer side. Other non-Bloom filter sub-plans are evaluated and added to the plan-list as usual.
    \end{itemize}
    \paragraph{Join Relation: $(t2,t3)$}
    \begin{itemize}
        \item $t2$ JOIN $t3$: Here, the required Bloom filter in sub-plan \linebreak $t3.\textrm{bf-subplan}_1$ cannot be resolved because the Bloom filter build relation, $t2$, is on the outer side. 
        \item $t3$ JOIN $t2$: Here, the required Bloom filter in sub-plan \linebreak $t3.\textrm{bf-subplan}_1$ cannot be resolved even though the Bloom filter build-column appears on the inner side; $t3.\textrm{bf-subplan}_1$ also requires $t1$ to appear on the inner side, i.e., it has property $\delta=\{t1,t2\}$.
    \end{itemize}
    \paragraph{Join Relation: $(t1,t2,t3)$}
    \begin{itemize}
        \item $(t1,t2)$ JOIN $t3$: Here, the sub-plan $t1.\textrm{bf-subplan}_0$ has already been resolved in relation $(t1,t2)$, 
        and the sub-plan $t3.\textrm{bf-subplan}_1$ cannot be resolved because it is on the inner side.
        \item $t3$ JOIN $(t1,t2)$: Here, the sub-plan $t3.\textrm{bf-subplan}_1$ can be resolved because both of its required relations, 
        $(t1,t2)$, appear on the inner side of the join. However, this joined sub-plan is rejected in our example because the estimated size of the Bloom filter is too large (Heuristic~\ref{h:BF_too_big}).
        \item $(t2,t3)$ JOIN $t1$: Here, the sub-plan $t1.\textrm{bf-subplan}_0$ cannot be resolved because the required Bloom filter build relation is on the outer side.
        \item $t1$ JOIN $(t2,t3)$: Here, the sub-plan $t1.\textrm{bf-subplan}_0$ can be resolved 
        because the required relation, $t2$, appears on the inner side of the join.
        We compute the cost of the corresponding hash join and allow 
        it to compete with the existing sub-plans in the plan-list for relation $(t1,t2,t3)$. 
        However, in our example, it is rejected from the plan-list because there is an existing plan that has lower cost. 
    \end{itemize}

    \noindent The winning plan in our running example is shown in Figure~\ref{fig:running_ex}. It is one that 
    applies a Bloom filter to $t1$ built from $t2$, then joins in $t3$. 

\end{example}

\subsection{Post-processing}

Our costing method is limited to a single select-project-join query block, 
but there are sometimes useful Bloom filters that can be pushed 
through a sub-query. For this reason, we retain the post-processing application 
of Bloom filters once the plan tree has been determined by BF-CBO. Bloom filters are 
added in places where either costing has determined they should be or where the 
pre-existing post-processing approach would have marked one. Post-processing repeats the 
assertion that the selectivity of the Bloom filter 
(ignoring false positives) be larger than a threshold and several other heuristics.

\subsection{Integration}

We integrate our BF-CBO method into \gauss{}~\cite{Huawei2024} (see also~\cite{Li2021,Memarzia2024,Li2024}). \gauss{} is a 
cloud-native distributed relational database system with a bottom-up 
query optimizer. Its query optimizer is extended from that of PostgreSQL~\cite{Postgres2024},
notably with added support for distribution across multiple nodes as well as 
symmetric multiprocessing (SMP).
We added Bloom filters to the \gauss{} optimizer as a planner post-process 
to serve as a baseline (\textbf{BF-Post}), and our two-phase bottom-up \textbf{BF-CBO}
approach is integrated into the cost-based optimization of a single 
query block (i.e., a single select-project-join block). The current work represents 
an initial implementation of BF-CBO that hasn't yet been fully tuned in terms of 
heuristics applied; we expect further improvements once this work is complete.

\subsection{Runtime Implementation}
\label{sec:runtime}

Our execution engine currently supports applying Bloom filters in an SMP,
 single-node deployment. In this setup, hash 
joins with a degree of parallelism (DOP) larger than one can be 
executed with various well-known streaming strategies. These streaming 
strategies influence how Bloom filters are used as follows:
\begin{enumerate}

    \item \emph{Broadcast join, build-side broadcast}. In this scenario, the build-side relation can originate 
    from a single thread and be broadcast to $n$ threads before computing the 
    hash join with the probe side's $n$ threads. In this case, we build only one  
    Bloom filter from one of the $n$ redundant hash tables on the build-side and use 
    it on the Bloom filter apply-side relation. 
    
    \item \emph{Broadcast join, probe-side broadcast}. In this scenario, the probe-side relation 
    originates from a single thread before being broadcast to perform 
    the hash join with $n$ threads on the build-side. In this case, the build side's
    $n$ threads are not redundant, so we must create individual Bloom filters 
    for each thread. We merge these Bloom filters by performing a union of their 
    bit vectors and apply the merged Bloom filter to the single-threaded apply-side.
    
    \item \emph{Partition join, partition-unaligned}. In this scenario, both the 
    build-side and the probe-side of the hash join are multithreaded and a 
    redistribution operation on either side may be necessary to shuffle the data by 
    grouping common values of a join column before $n$ partial hash joins are 
    computed independently on each group of values. In this case, we build $n$ 
    partial Bloom filters on the build-side, one for each partition of the hash join. 
    A Bloom filter can, in general, be applied 
    to a relation that is under 
    some intermediate nodes on the probe-side, and the partitioning of that relation 
    is not necessarily the same as the partitioning of the hash join in which 
    the Bloom filter is built. When partitioning is \textit{unaligned} like this, 
    we can use the value of the Bloom filter 
    partitioning column for distributed lookup of which Bloom filter partition 
    to use, provided the partitioning column is available on the apply-side relation.
    When unavailable, we can use the bit vector merging strategy described for Broadcast join.

    \item \emph{Partition join, partition-aligned}. This scenario is also a partition 
    join, but in this case, the partitioning of the relation to which the Bloom filter 
    is applied is aligned with the partitioning of the hash join build-side. Here, 
    we build $n$ partial Bloom filters on the build-side in the same way as the \textit{partition unaligned} case. 
    On the apply-side of the Bloom filter, for each partition of a relation, we simply 
    apply the appropriate Bloom filter partition. 

\end{enumerate}

While it is possible to account for these different streaming strategies in 
the Bloom filter cost model, we did not do so for the results in this paper. 
Table scans wait for all Bloom filter partitions to become available before 
scanning can proceed, regardless of streaming strategy.

\subsection{Bloom filter limiting heuristics}
\label{sec:heuristics}

Throughout the description of our two-phase bottom-up method, we described 
several heuristics we applied to limit the search space of evaluating 
Bloom filter sub-plans or to improve expected efficiency. They were 
described in the context of where 
we implemented them, but we list them here as a summary for the reader.

\begin{itemize}
    \item \heuristic{Bloom filter candidates are only applied on the larger relation for each hashable join clause (Section~\ref{sec:methods_marking_bfcs}). \label{h:larger_rel}}
    \item \heuristic{Bloom filter candidates are only applied on relations whose estimated cardinality surpasses a threshold (Section~\ref{sec:methods_marking_bfcs}). \label{h:apply_threshold}}
    \item \heuristic{Bloom filters cannot be applied to foreign keys joining with \textit{lossless} primary keys (Section~\ref{sec:methods_phase1}). \label{h:fk_pk}}
    \item \heuristic{All Bloom filter candidates that can be applied on a relation must be applied simultaneously when creating scan sub-plans (Section~\ref{sec:methods_costing_subplans}). \label{h:all_BFCs}}
    \item \heuristic{If the expected size of the Bloom filter is beyond a threshold, a Bloom filter is not created (Section~\ref{sec:methods_costing_subplans}). \label{h:BF_too_big}}
    \item \heuristic{Bloom filters whose estimated selectivity is below a threshold are removed (Section~\ref{sec:methods_costing_subplans}). \label{h:selec_threshold}}
\end{itemize}

Several other heuristics could be applied to limit the search space of 
evaluating Bloom filter sub-plans, which we did not implement in our main results (Section~\ref{sec:results}). For example:

\begin{itemize}
    \item \heuristic{During planning, if a relation has too many Bloom filter sub-plans, only the one with the fewest estimated rows should be kept. \label{h:limit_search_space}} Ties may be broken by keeping only the sub-plan with the lowest total cost. This heuristic should limit the search space of BF-CBO. We explore its effect in Section~\ref{sec:limit_search_space}.
    \item \heuristic{If the total join-input cardinality during bottom-up phase 1 is below a threshold, adding Bloom filter candidates should be skipped. \label{h:total_join_threshold}} 
    With no Bloom filter candidates, the planning search space will not be expanded, so the second bottom-up phase would revert to normal CBO.
    This heuristic is meant to differentiate quick transactional queries, where additional optimization time is not necessary, 
    from long-running analytical queries, where the additional time spent considering Bloom filters during planning can make a big difference. 
    The total join-input cardinality can be computed as the cumulative sum of the cardinality of join inputs, for all joins considered during bottom-up phase 1.
    The maximum join-input size could be an additional signal to decide if Bloom filter candidates should be skipped.
    \item \heuristic{Allow Bloom filter candidates to be applied to both relations in a join clause, but keep only the $\delta$s that are smaller than the apply-side relation}. 
    This is a slightly more permissive alternative to Heuristic~\ref{h:larger_rel} that will consider Bloom filters for relations 
    that are larger than the build-side relation, for any combination of joined relations that make up the build side. This allows 
    a Bloom filter candidate to be applied to the smaller base relation of a join-clause pair, but only for cases where an intermediate join will 
    reduce the size of the larger base relation of the join-clause pair.
\end{itemize}

\section{Experimental analysis}
\label{sec:experiments}

\subsection{Dataset and environment}
\label{sec:tpch_queries}

We ran our analysis on a TPC-H dataset of scale factor 100 (approximately 100 GB). 
Each of the 22 TPC-H queries was run five times, with the average of the last 
four presented here to represent performance after data had been loaded
into memory. The dataset tables were stored in a columnar format, 
range-partitioned by date, and foreign key 
constraints were added in compliance with TPC-H documentation. We ran all 
queries with a DOP of 48 as our experiments were performed on an x86 server 
with 48 CPUs and 503 GB memory. Several queries were run with query-specific 
database configuration parameters as other areas of our optimizer are 
actively being refined; we held these configuration parameters fixed between 
the baselines and BF-CBO for a fair comparison.

In our experiments, the selectivity threshold was set to 
$\frac{2}{3}$, so that Bloom filter 
candidates were kept only if they were expected to filter out at least 
$\frac{1}{3}$ of the rows (Heuristic~\ref{h:selec_threshold}). We only marked Bloom filter candidates if 
the number of rows in the table they were applied to was greater than 10 thousand (Heuristic~\ref{h:apply_threshold}). 
Bloom filters were considered too big if the estimated upper-bound number 
of distinct values on the build side was beyond 2 million (Heuristic~\ref{h:BF_too_big}).

\subsection{Results}
\label{sec:results}

\begin{figure}[tb]
    \centerline{\includegraphics[width=0.95\columnwidth]{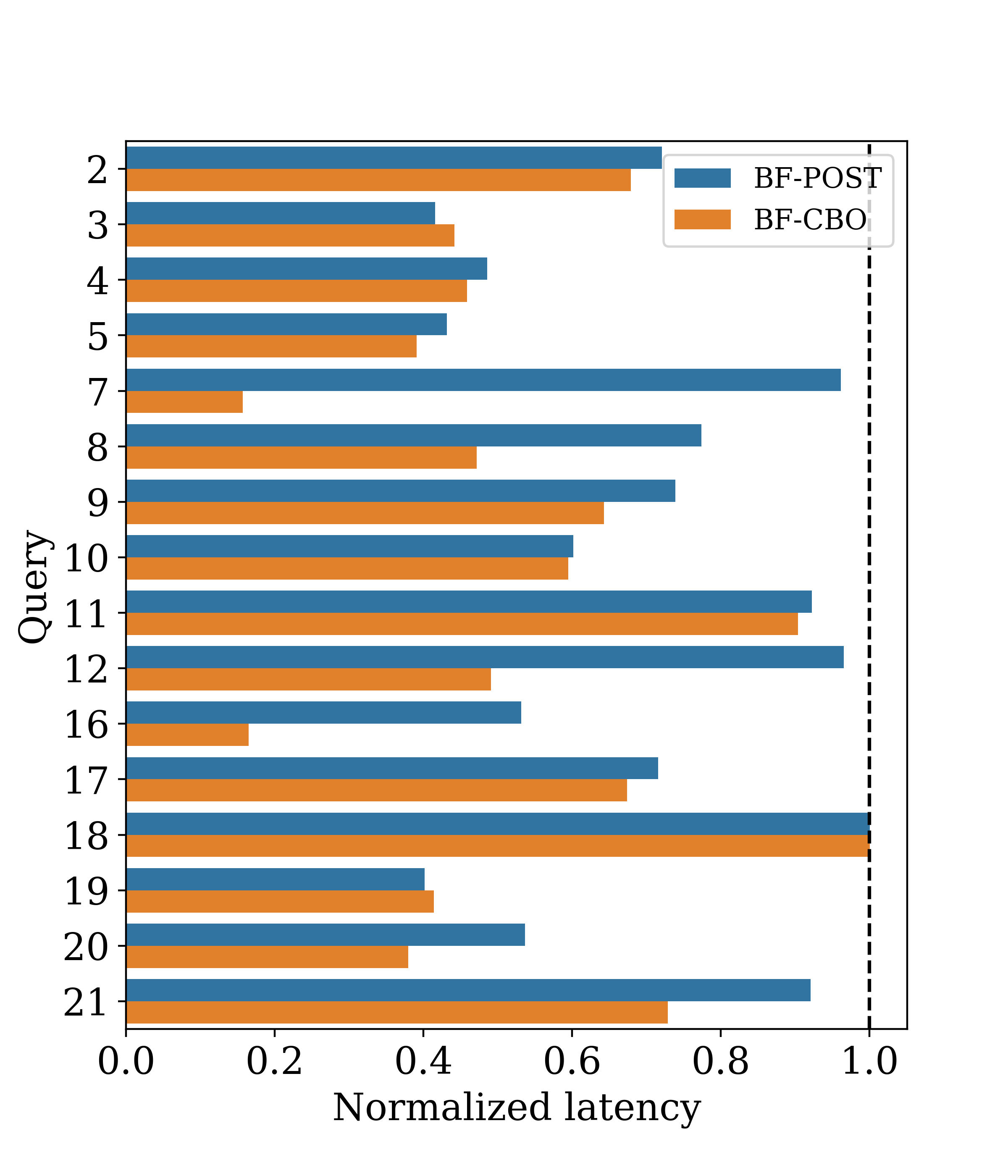}}
    \caption{
        Latencies for TPC-H queries are shown normalized to each query without any Bloom 
        filters applied (No BF, shown as dashed line). Adding Bloom 
        filters during plan post-processing (BF-Post, shown in blue) reduces 
        query latency by 28.8\%. Including Bloom filters during bottom-up 
        cost-based optimization (BF-CBO, shown in orange)
        improves query latency by a further 32.8\% relative to BF-Post. TPC-H queries that did 
        not apply any Bloom filters are omitted.
    }
    \label{fig:tpch_results}
    \Description{A bar graph compares 3 methods for each tpch query.}
\end{figure}

The latencies of the TPC-H queries are shown in Figure~\ref{fig:tpch_results} 
with more details in Table~\ref{tab:tpch_results}. Single table queries (Q1 and Q6),
as well as queries that did not produce Bloom filters in any scenario (Q13-15,22), 
are omitted from the analysis. The latencies are normalized to the latency
of running the query without Bloom filters enabled (\textbf{No BF}). Table~\ref{tab:tpch_results} 
also shows the percent reduction in query latency (\% $\downarrow$) of BF-CBO compared 
to BF-Post as well as the absolute latencies of plan optimization
for both BF-CBO and BF-Post. Note that planner runtime 
is included in the measurement of absolute query latency before 
normalization. Query numbers (Q\#) where BF-CBO 
selected a different plan than BF-Post are shown italicized in \textcolor{red}{\textit{\textbf{red}}}.

\begin{table}[tb]
    \caption{TPC-H query latencies}
    \centering
    \begin{tabular}{l|rr|r|rr}
        \toprule
                       & \multicolumn{3}{|c}{Normalized query latency} & \multicolumn{2}{|c}{planner latency (ms)} \\
        \textbf{Q\#} & \textbf{BF-Post} & \textbf{BF-CBO} & \textbf{\% $\downarrow$} & \textbf{BF-Post} & \textbf{BF-CBO} \\
        \midrule
        {\color[HTML]{FF0000} \textit{\textbf{2}}}  & 0.721 & 0.679 & 5.9  & 18.4  & 24.9  \\ \hline
        3                                           & 0.416 & 0.442 & -6.1 & 1.8   & 2.1   \\ \hline
        4                                           & 0.486 & 0.459 & 5.6  & 0.7   & 0.7   \\ \hline
        {\color[HTML]{FF0000} \textit{\textbf{5}}}  & 0.432 & 0.391 & 9.5  & 22.8  & 41.0  \\ \hline
        {\color[HTML]{FF0000} \textit{\textbf{7}}}  & 0.962 & 0.157 & 83.7 & 17.6  & 28.3  \\ \hline
        {\color[HTML]{FF0000} \textit{\textbf{8}}}  & 0.774 & 0.472 & 38.9 & 61.6  & 166.3 \\ \hline
        {\color[HTML]{FF0000} \textit{\textbf{9}}}  & 0.739 & 0.643 & 12.9 & 81.6  & 155.2 \\ \hline
        10                                          & 0.602 & 0.595 & 1.2  & 1.7   & 2.0   \\ \hline
        {\color[HTML]{FF0000} \textit{\textbf{11}}} & 0.923 & 0.904 & 2.0  & 8.0   & 11.1  \\ \hline
        {\color[HTML]{FF0000} \textit{\textbf{12}}} & 0.966 & 0.491 & 49.2 & 0.8   & 0.8   \\ \hline
        {\color[HTML]{FF0000} \textit{\textbf{16}}} & 0.532 & 0.165 & 69.1 & 3.3   & 4.2   \\ \hline
        17                                          & 0.716 & 0.674 & 5.9  & 1.8   & 2.0   \\ \hline
        {\color[HTML]{FF0000} \textit{\textbf{18}}} & 1.001 & 1.001 & 0.0  & 5.5   & 9.4   \\ \hline
        19                                          & 0.402 & 0.414 & -3.0 & 1.2   & 1.2   \\ \hline
        {\color[HTML]{FF0000} \textit{\textbf{20}}} & 0.537 & 0.380 & 29.3 & 6.0   & 7.7   \\ \hline
        {\color[HTML]{FF0000} \textit{\textbf{21}}} & 0.921 & 0.729 & 20.8 & 21.5  & 83.8  \\ \hline
        total                                       & 0.712 & 0.478 & 32.8 & 254.3 & 540.7 \\ 
        \bottomrule
    \end{tabular}
    \label{tab:tpch_results}
\end{table}

Across all analyzed queries, including Bloom filters in plan post-processing (BF-Post) 
reduced the runtime by 28.8\%, and including Bloom filters during 
bottom-up CBO reduced the runtime by 52.2\%, relative to no Bloom filters at all.
The addition of BF-CBO led to a 32.8\% reduction in runtime compared to BF-Post, so there is
a significant benefit to including the effect of Bloom filters during CBO
rather than simply as a post process.
Several queries had a large reduction in latency, 
such as Q7, Q8, Q12, Q16, Q20, and Q21. We will examine the query plans from 
some of these queries in subsequent sections to explain these improvements. 

Table~\ref{tab:tpch_results} also shows that there is some overhead in
planner runtime with BF-CBO compared to BF-Post. To plan all 
the queries, BF-CBO took 540.7 ms while BF-Post took 254.3 ms. Many queries
showed negligible overhead when using BF-CBO, but some queries, like Q8 and Q9, had 
large increases in planner latency. Increased planner runtime is expected with BF-CBO as there are more 
sub-plan combinations to search, but end-to-end, we see a large improvement 
in query latencies. The trade-off between increased planner runtime and query plan 
improvement will depend on the context, with BF-CBO being more appropriate 
for long-running analytical queries rather than quick transactional queries.

As incorporating Bloom filters during query optimization will adjust
the cardinality of scan nodes where the Bloom filter is applied, we 
expected an improvement in cardinality estimation as well. We found that
BF-CBO had a mean absolute error (MAE) of $5.3e^{6}$ for the cardinality estimates of all 
intermediate plan nodes, compared to $2.5e^{7}$ for BF-Post, a 78.8\% improvement.
In fact, it follows that improving the cardinality estimate of Bloom filter 
table scans \textit{enables} the improved query plans in this paper.

\subsection{Query Plan Analysis}

\begin{figure*}[t!]

    \sbox\twosubbox{%
    \resizebox{\dimexpr.9\textwidth-1em}{!}{%
        \includegraphics[height=3cm]{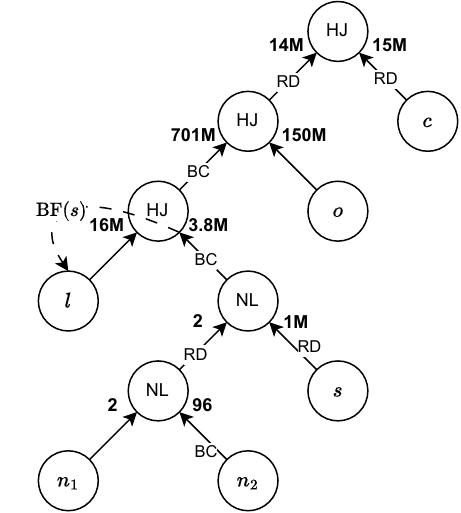}%
        \includegraphics[height=3cm]{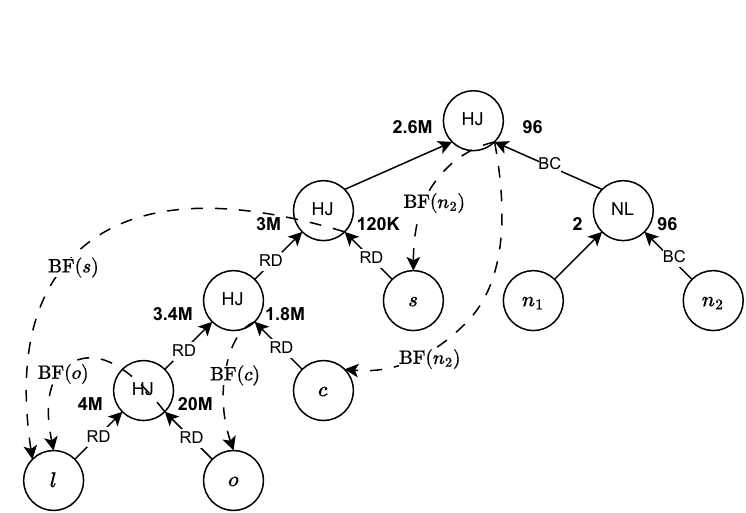}%
    }%
    }
    \setlength{\twosubht}{\ht\twosubbox}

    \centering
    \subcaptionbox{BF-Post\label{fig:q7a}}{%
        \includegraphics[height=\twosubht]{q7_bfp1_round}%
    }\quad
    \subcaptionbox{BF-CBO\label{fig:q7b}}{%
        \includegraphics[height=\twosubht]{q7_bfp2_round}%
    }
    \caption{
        Join orders for TPC-H query 7. Observed input row counts are shown 
        in bold on the left (outer/probe) and right (inner/build) sides 
        of each join. Streaming across threads is denoted by RD (redistribution) 
        or BC (broadcast). Post-processing application 
        of Bloom filters (panel \subref{fig:q7a}) applies a single 
        Bloom filter on the relation $l$:\textit{lineitem} 
        built with respect to the relation $s$:\textit{supplier} (black dashed arrow labeled BF($s$)). 
        BF-CBO (panel \subref{fig:q7b})
        changes the join order so that five Bloom filters can be applied, 
        reducing the input row counts to many joins and improving the 
        query latency by 83.7\%.
    }
    \label{fig:q7}
    \Description{Two subfigures are shown each depicting the join tree for tpch query 7 obtained through different optimization methods.}
\end{figure*}

We showed in Figure~\ref{fig:q12} that BF-CBO allowed for the selection 
of a plan that positioned \textit{orders} on the probe side of the hash join, 
provided that a Bloom filter would be applied during its scan. The reason 
this plan could be selected is that BF-CBO includes a sub-plan with a 
Bloom filter applied to \textit{orders}, reducing the row estimate of 
\textit{orders} from 150 million to 6.4 million rows, which in turn reduces the estimated
cost of performing the hash join.

Another example is shown in Figure~\ref{fig:q7} for TPC-H query 7, which 
shows the join order for its \textit{FROM} clause, which appears in Listing~\ref{lst:tpch_q7}. In this case, 
BF-CBO allows for a different join order that enables 
five Bloom filters to be applied instead of just one. BF-CBO uses a Bloom filter
to significantly reduce the size of several large tables in query 7.  
It applies two Bloom filters to the \textit{lineitem} table, reducing its row count 
to 4 million (relative to 16 million in BF-Post), and it applies a Bloom filter to 
the \textit{orders} table, reducing its row count from 150 million to 20 million.
The \textit{customer} and \textit{supplier} tables are also significantly 
reduced using Bloom filters. These reductions in row count mean that 
joins throughout the query are faster and query performance is better. 

One 
of the reasons this join order can achieve these reductions in row 
count is its effective predicate transfer. Note that in BF-Post (panel~\subref{fig:q7a}), 
there is no Bloom filter applied to the \textit{lineitem} table originating from \textit{orders}---the 
reason for its absence is that the join clause between these two 
relations (\textit{o.orderkey} = \textit{l.orderkey}) consists of a foreign key column (of \textit{lineitem}) referencing an \textit{unfiltered} 
primary key column (of \textit{orders}). As explained earlier, in this scenario,
a Bloom filter will not filter any rows of the foreign key column. However,
when including Bloom filters in bottom-up CBO (panel~\subref{fig:q7a}), the planner 
has information that the Bloom 
filter applied to \textit{orders} (i.e., BF$(c)$) filters out some of those primary keys,
enabling the additional Bloom filter on \textit{lineitem} (i.e., BF$(o)$). Similarly, the 
Bloom filter on \textit{orders} (BF$(c)$) is only enabled by BF-CBO because
the \textit{customer} relation is pre-filtered by another Bloom filter 
based on \textit{n2.nationkey}. So, we have effective predicate transfer 
of the \textit{nation} predicates---they reduce the size of the \textit{customer} 
table, which in turn allows a Bloom filter to reduce the size of the \textit{orders} 
table, which in turn allows a Bloom filter to reduce the size of the \textit{lineitem} 
table. 

Another reason BF-CBO performs effectively is that it places Bloom filters in 
such a way that they can be pushed down through a join. By crossing 
an intermediate join, each Bloom filter effectively reduces the input 
of multiple joins, magnifying its effect.

\begin{lstlisting}[language=SQL, caption={TPC-H Q7 FROM Clause}, label={lst:tpch_q7}]
select
    n1.n_name as supp_nation,
    n2.n_name as cust_nation,
    extract(year from l_shipdate)
        as l_year,
    l_extendedprice * (1 - l_discount) 
        as volume
from
    supplier,
    lineitem,
    orders,
    customer,
    nation n1,
    nation n2
where
    s_suppkey = l_suppkey
    and o_orderkey = l_orderkey
    and c_custkey = o_custkey
    and s_nationkey = n1.n_nationkey
    and c_nationkey = n2.n_nationkey
    and ((n1.n_name = 'FRANCE' 
          and n2.n_name = 'GERMANY') or 
         (n1.n_name = 'GERMANY' 
          and n2.n_name = 'FRANCE'))
    and l_shipdate between 
        date '1995-01-01' 
        and date '1996-12-31'
\end{lstlisting}

Readers will also observe from Table~\ref{tab:tpch_results} that adding 
Bloom filters to query 18 did not improve its runtime, even for BF-Post.
The runtime of query 18 is dominated by a sub-query (not shown) to which no Bloom 
filter is applied, so adding Bloom filters to other table scans in the 
query, built on the output of this sub-query, did not result in improved latency overall. 
Part of the reason latency is not improved may 
be that, in our implementation, table scans wait for any required Bloom filters to be fully 
built before the scan can proceed. So when the query runs without any Bloom filters, 
those scans may have been able to start in parallel with the sub-query, 
instead of waiting for it to complete. An alternative implementation 
could eagerly scan batches of data before the Bloom filter is fully built to take 
advantage of parallel processing, then once ready, switch to using the Bloom filter 
for any remaining batches to be scanned. However, we believe that it is usually better to wait
for the Bloom filter before starting the scan, 
because down-stream operators will benefit from reduced rows compared to eager
scanning. When using BF-CBO, the entire down-stream query plan assumes that 
all Bloom filters will be fully utilized---violating this assumption may be
detrimental.

\subsection{Limiting Bloom filter sub-plans}
\label{sec:limit_search_space}

In this section, we analyze the effect of enabling Heuristic~\ref{h:limit_search_space}, which limits the 
search space of Bloom filter sub-plans during bottom-up optimization. 
Specifically, if any given relation has too many Bloom filter sub-plans 
(more than four in our experiments) during bottom-up optimization, we prune those sub-plans down to only 
one---we keep the one with the fewest expected rows (or the lowest cost, if rows are equal).
By enabling this heuristic, we expect planning to be quicker, but  
with some opportunity lost in finding the best query plan. The results for this 
restriction are shown in Table~\ref{tab:tpch_results_nopathlimit}. 
The queries where BF-CBO resulted in different query plans than in Table~\ref{tab:tpch_results}
are shown italicized in {\color[HTML]{00B050} \textit{\textbf{green}}}. The columns 
for BF-Post are identical to those in Table~\ref{tab:tpch_results}, but 
are repeated here for convenience.

\begin{table}[tbh]
    \caption{TPC-H query latencies, Heuristic~\ref{h:limit_search_space} enabled}
    \centering
    \begin{tabular}{l|rr|r|rr}
        \toprule
                       & \multicolumn{3}{|c}{Normalized query latency} & \multicolumn{2}{|c}{planner latency (ms)} \\
        \textbf{Q\#} & \textbf{BF-Post} & \textbf{BF-CBO} & \textbf{\% $\downarrow$} & \textbf{BF-Post} & \textbf{BF-CBO} \\
        \midrule
        {\color[HTML]{00B050} \textit{\textbf{2}}} & 0.721 & 0.741 & -2.7 & 18.4  & 21.3  \\ \hline
        3                                          & 0.416 & 0.442 & -6.2 & 1.8   & 2.0   \\ \hline
        4                                          & 0.486 & 0.457 & 5.9  & 0.7   & 0.7   \\ \hline
        5                                          & 0.432 & 0.379 & 12.3 & 22.8  & 32.6  \\ \hline
        7                                          & 0.962 & 0.148 & 84.6 & 17.6  & 22.6  \\ \hline
        {\color[HTML]{00B050} \textit{\textbf{8}}} & 0.774 & 0.805 & -4.0 & 61.6  & 107.2 \\ \hline
        9                                          & 0.739 & 0.603 & 18.4 & 81.6  & 153.1 \\ \hline
        10                                         & 0.602 & 0.609 & -1.2 & 1.7   & 2.0   \\ \hline
        11                                         & 0.923 & 0.842 & 8.8  & 8.0   & 11.1  \\ \hline
        12                                         & 0.966 & 0.483 & 49.9 & 0.8   & 0.8   \\ \hline
        16                                         & 0.532 & 0.170 & 68.1 & 3.3   & 4.3   \\ \hline
        17                                         & 0.716 & 0.722 & -0.8 & 1.8   & 2.1   \\ \hline
        18                                         & 1.001 & 0.996 & 0.5  & 5.5   & 8.9   \\ \hline
        19                                         & 0.402 & 0.391 & 2.6  & 1.2   & 1.2   \\ \hline
        20                                         & 0.537 & 0.391 & 27.2 & 6.0   & 7.5   \\ \hline
        21                                         & 0.921 & 0.711 & 22.8 & 21.5  & 44.4  \\ \hline
        total                                      & 0.712 & 0.488 & 31.4 & 254.3 & 421.9 \\ 
        \bottomrule
    \end{tabular}
    \label{tab:tpch_results_nopathlimit}
\end{table}

The first notable difference when Heuristic~\ref{h:limit_search_space} is enabled is that planning latencies are shorter. In total, 
the planning time of all queries was 421.9 ms compared to 540.7 ms with 
Heuristic~\ref{h:limit_search_space} disabled. For queries Q8 and Q21, in particular, 
we save considerable time planning when we limit the search space by enabling Heuristic~\ref{h:limit_search_space}.
However, the query plan for Q8 is worse with the search space limited, 
and the query runtime now degrades by 4\% compared to BF-Post.

There is a trade-off between planning latency and finding the best query plan.
By limiting the search space through imposing Heuristic~\ref{h:limit_search_space} 
we observe faster query planning latencies, but overall query latency
is slightly degraded (a 31.4\% reduction in latency over BF-Post compared to 32.8\%),
indicating that for this dataset, it is still worthwhile 
to explore more paths. As such, our heuristics may require further tuning.

There are two potential explanations for the worse result observed in Q8 
when search space is limited. First, because we apply heuristics, we are 
removing some Bloom filter sub-plans from being considered. It is possible 
that the best plan appears in these removed Bloom filter sub-plans, but 
BF-CBO chooses a different plan because the cost of other sub-plans has 
been lowered by Bloom filters. BF-Post may 
arrive at the best plan by chance, as Bloom filters are not considered 
during planning. Second, the worse result could be due to imperfect 
cardinality estimations or an imperfect cost model.  
Our method can be thought of as improving the estimated cardinality of base tables 
to which Bloom filters are applied; but it still makes use of pre-existing 
methods for estimating join cardinality and a pre-existing cost model. An
imperfect cost model can sometimes lead to worse query plans, even with 
a better cardinality estimate.

\section{Conclusion and future work}
\label{sec:conclusion}

There are several promising avenues for future work. First, support could 
be extended to multi-node deployments. In multi-node deployments, the 
Bloom filters may need to be transferred and merged across nodes. This extra 
streaming will likely affect Bloom filter performance and should be 
factored into the cost model. Second, support for multi-column Bloom filters
could be added. In this work, we supported single-column Bloom filters: when there were 
multiple join columns, we planned for and built individual Bloom filters on
each column. Future work should explore how to include multi-column Bloom filters 
in cost-based optimization, which may have added benefit. Third, the runtime execution
could actively monitor Bloom filter bit vector saturation to adapt to poor estimation 
of the number of distinct entries. If the number of distinct entries is underestimated,
the Bloom filter may not filter any rows, in which case it is not worthwhile 
to send it to the probe side. Finally, further work may be needed to tune the heuristics 
we applied in this paper to ensure the correct trade-off between 
planner latency and query plan quality.

The favorable query plans presented in this paper are only made 
possible by including Bloom filters directly in cost-based optimization. 
Our paper describes the challenges of incorporating Bloom filter cost into 
a bottom-up optimizer. In particular,  
we showed that pruning Bloom-filter sub-plans is not possible until 
the full set of relations appearing on the build side of the hash join 
that creates the Bloom filter is known, at which point those sub-plans 
can be fully costed. We proposed an efficient two-phase bottom-up 
approach that defers planning and costing of sub-plans until pruning 
is possible and avoids an unnecessary expansion of the search space. Despite this 
efficient two-phase approach, the search space when including Bloom filter 
sub-plans is increased. We applied several search-space-limiting heuristics 
to handle this increase. Our method obtained a 32.8\% reduction in query latency 
for TPC-H queries that involved Bloom filters, and we demonstrated 
the reasons for that improvement.

\balance
\bibliographystyle{ACM-Reference-Format}
\bibliography{bf_paper}

\end{document}